\documentclass[letter]{aa}

\usepackage{graphicx}

\usepackage{txfonts}

\usepackage{natbib}

\bibpunct{(}{)}{;}{a}{}{,}

\newcommand{\tpe}[1]{\zeta_\mathrm{pe#1}}

\begin{document}
   \title{Observation of a short-lived pattern in the solar chromosphere}
   \author{F. W\"oger
          \and
          S. Wedemeyer-B\"ohm
          \and
          W. Schmidt
          \and
          O. von der L\"uhe
          }
   \offprints{F. W\"oger, woeger@kis.uni-freiburg.de }
   \institute{Kiepenheuer-Institut f\"ur Sonnenphysik,
              Sch\"oneckstra{\ss}e 6, D-79104 Freiburg, Germany\\
              \email{[woeger,wedemeyer,wolfgang,ovdluhe]@kis.uni-freiburg.de}
             }
   \date{Received August 11, 2006; accepted September 13, 2006}

\abstract{}
{In this work we investigate the dynamic behavior of inter-network
  regions of the solar chromosphere.}
{We observed the chromosphere of the quiet Sun using a narrow-band
Lyot filter centered at the \mbox{{\ion{Ca}{ii}\,K$_\mathrm{2v}$}}
emission peak with a bandpass of 0.3\,\AA. We achieved a spatial
resolution of on average 0\farcs 7 at a cadence of 10\,s.}
{In the inter-network we find a mesh-like pattern
that features bright grains at the vertices. The pattern has
a typical spatial scale of 1\farcs 95 and a mean evolution time scale of
\mbox{53\,s} with a standard deviation of \mbox{10\,s}. A comparison
of our results with a recent three-dimensional radiation
hydrodynamical model implies that the observed pattern is of
chromospheric origin. The measured time scales are not compatible
with those of reversed granulation in the photosphere although the
appearance is similar. A direct comparison between network and
inter-network structure shows that their typical time scales differ
by at least a factor of two.}
{The existence of a rapidly evolving small-scale pattern in the inter-network regions supports the picture of the lower chromosphere as a highly dynamical and intermittent phenomenon.}


\maketitle


\section{Introduction}

Deducing the thermal structure of the solar chromosphere is complicated by the limited number of diagnostics and the difficulties to interpret them.
An example are the core regions of the calcium resonance lines (\mbox{\ion{Ca}{ii}\,H and K}) that are commonly assumed to originate from the chromosphere \citep[cf.][]{val81}.
Unfortunately the simplifying assumption of local thermodynamic equilibrium (LTE) is not valid for the emission reversal peaks and the cores of these lines.
Rather, non-local effects must be taken into account for a realistic description of chromospheric plasma and radiation field.
This is not only a problem for the interpretation of observations of these spectral features but also for the numerical modeling of the chromosphere.
The recent three-dimensional radiation hydrodynamics model by \citet[][hereafter W04]{wedemeyer04a} shows a mesh-like pattern at chromospheric heights in non-magnetic inter-network regions, consisting of cool regions and hot filaments that are a product of propagation and interaction of oblique shock waves and adiabatic expansion of post-shock regions.
The authors argue that the amplitude of the temperature variations in the model chromosphere is somewhat uncertain due to the too simple radiative transfer (grey, LTE).
But the pattern is nevertheless representative as it is a product of acoustic waves that are excited in the lower well-modeled layers.
The spatial scales of the pattern are comparable to the underlying granulation but the dynamical time scales are much shorter.
Mapping the chromospheric small-scale structure of inter-network regions thus requires both high spatial and temporal resolution.
This might be the reason why most earlier observations failed to detect this phenomenon (Table~\ref{tbl:cadata}).

Here we present new observations taken with a Lyot filter of very narrow transmitted wavelength range and a spatial resolution high enough to resolve the chromospheric small-scale pattern of inter-network regions.

In Sect.~\ref{sec:obs} we describe the observations and data reduction, followed by results and discussion in Sects.~\ref{sec:result} and \ref{sec:discuss}, respectively.


\begin{figure*}[htp]
 \centering
 \includegraphics[width=18cm]{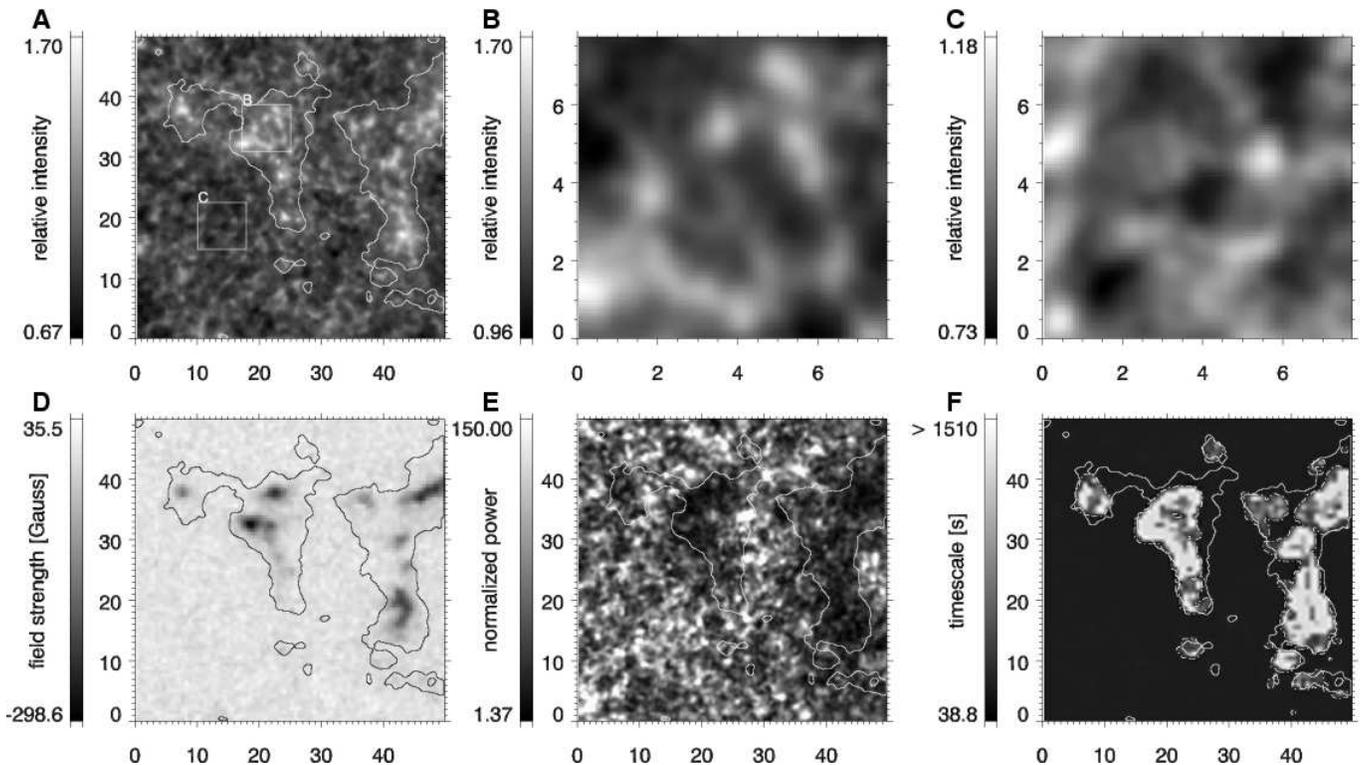}
 \caption{(A) Calcium image with superimposed network mask and closeup regions (numbered squares), (B) the closeups of the network region and (C) the inter-network region, (D) the MDI magnetogram, (E) a  2-D power map of the temporal frequency band at \mbox{5.5\,mHz} with a width of \mbox{0.9\,mHz} and (F) the map of \mbox{1/e} time scales. All axes are in arcseconds.
 }
 \label{fig:caimage}
\end{figure*}

\begin{table}[pb]
  \centering
  \caption{Recent observations of the \ion{Ca}{ii} resonance lines. The properties of the data show that the presented data set is unique regarding the combination of spectral and spatial resolution (\mbox{$\Delta\lambda$} is FWHM).
  }
  \begin{tabular}{lrrrr}
   \hline
   Author & $\lambda$ & $\Delta\lambda$ & scale        & $\Delta$t\\
          & $[\AA]$   & $[\AA]$         & [\arcsec/px] & [s]    \\
   \hline
   this work & 3933.5 & 0.30 & 0.1485 & 10\\
   \cite{rutten04dot2} & 3968.8 & 1.35 & 0.071 & 30\\
   \cite{tritschler02} & 3933.2 & 0.60 & 0.350 & 6\\
   \cite{depontieu02} & 3933.0 & 3.00 & 0.083 & $\sim$20\\
   \cite{kneer93} & 3933.0 & 0.60 & 0.720 & 15\\
   \hline
  \end{tabular}
  \label{tbl:cadata}
\end{table}


\section{Observation and data reduction}

\label{sec:obs}

\begin{figure}[tp]
  \centering
  \includegraphics[width=8.5cm]{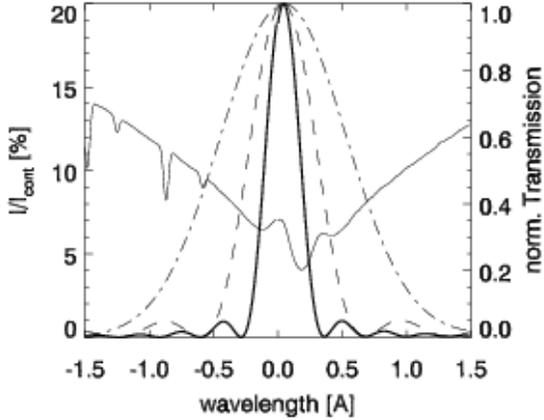}
  \caption{The calcium line around \mbox{3933.5\,\AA} \citep{neckel99}
    (thin solid) plotted together with theoretical Lyot filter
    transmission curves with \mbox{0.3\,\AA} (thick solid),
    \mbox{0.6\,\AA} (dashed) and a Gaussian with \mbox{1.350\,\AA}
    (dot-dashed) FWHM (see Table \ref{tbl:cadata}).  }
  \label{fig:lyot}
\end{figure}

A high-resolution image sequence was obtained with the German Vacuum Tower Telescope (VTT) at the Observatorio del Teide on Tenerife, Spain, during UT~8:25--10:21 on April~18, 2005.
The telescope was pointed at a quiet area near disk-center.
The field of view is \mbox{49\arcsec $\times$\,49\arcsec} showing chromospheric network and inter-network.
The data set consists of images observed in the G-Band \mbox{(4305\,{\AA})} using an interference filter with a FWHM of \mbox{10\,\AA}, and, using Lyot filters, in \mbox{\ion{Ca}{ii}\,K} \mbox{(3933\,{\AA})} and H$\alpha$ \mbox{(6563\,{\AA})} with FWHMs of \mbox{0.3\,\AA} and \mbox{0.25\,\AA}, taken simultaneously.
In this paper we only consider the results obtained from the calcium filtergrams.
A sample image together with closeups of a network and an inter-network region is shown in Fig.~\ref{fig:caimage}.
The filter curve of the Ca filter is displayed in Fig.~\ref{fig:lyot} along with filter curves of other observations for comparison.
The image sequence was taken at a cadence of \mbox{10\,s} and a pixel scale of \mbox{0\farcs 1485/px}.
There was a need for telescope adjustments at UT~9:00, separating the sequence into two parts with a gap of 2~min.
The transmission of the Lyot filter is rather poor, thus the light level of the \mbox{\ion{Ca}{ii}\,K} line core -- being by itself less than 5\% of the continuum intensity --  was reduced even further, leading to an exposure time of \mbox{2\,s}.

All images were corrected for hot pixels and calibrated using the
standard procedure of flat-fielding.  Additionally, the sequence was
corrected for the change in mean intensity in each image due to the
increasing solar elevation with a fitted sine curve and scaled to the
mean intensity of the time series.  Finally, the noise in each image
was reduced using a Wiener Filter.
From the cutoff of the
azimuthally integrated spatial power spectrum we estimate a spatial
resolution of on average \mbox{0\farcs 7} or better which was achieved by the
fair to good seeing (\mbox{$r_0$\,$\approx$13\,cm}) in combination
with the Kiepenheuer Adaptive Optics System (KAOS).


\section{Analysis and results}

\label{sec:result}
\begin{figure}[tp]
  \centering
  \includegraphics[width=8.5cm]{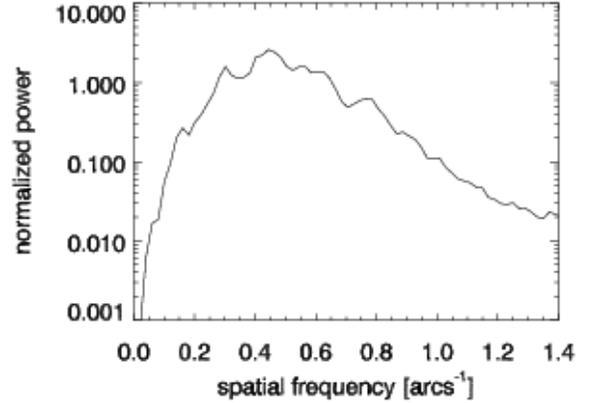}
  \caption{One-dimensional power spectrum of the inter-network
    pattern.  }
  \label{fig:sps}
\end{figure}

In our analysis we focus on the second half of the image series with a
length of \mbox{75\,min}.  Network regions embody chains of persistent
bright points (Fig.~\ref{fig:caimage}B).  Inter-network regions show a
highly dynamic behavior with small-scale, cell-like structures
(Fig.~\ref{fig:caimage}C)
and small brightenings at the vertices that resemble bright points or grains
\citep[see, e.g.,][]{lites99}.
We computed the spatial power spectrum of the inter-network pattern.
In order to enhance the small-scale nature of the pattern of interest,
an unsharp-masked image was subtracted from the data.  The
two-dimensional power spectrum of the resulting image was then
azimuthally integrated.  The one-dimensional power spectrum is shown
in Fig.~\ref{fig:sps}.  The barycenter of power is at spatial
frequency \mbox{$f_h=0.512$ arcs$^{-1}$} corresponding to
\mbox{1\farcs 95}.
The bright grains have typically a size of 1-2\,\arcsec
(see Fig.~\ref{fig:caimage}C).

To demonstrate the dynamical behavior in further detail, we calculated
the two-dimensional autocorrelation function for subfields of
\mbox{1\farcs 485\,$\times$\,1\farcs 485} which overlap by half of
their field size in both directions.  The evolution time scale
$\tpe{}$ is defined as the time at which the autocorrelation has
dropped to a value of \mbox{1/e}.  Finding the evolution time scale
for each subfield separately results in an array of
\mbox{66\,$\times$\,66} values over the field of view.  The map in
Fig.~\ref{fig:caimage}F shows the results for the individual
subfields.  Figure~\ref{fig:time_int} shows a plot of time scale
versus mean intensity for all subfields.  In order to persue a more
detailed analysis of the different behavior of network and
inter-network, criteria need to be established which distinguish these
two regions.

As a first step, the lower mean brightness of the inter-network
pattern was used to create a mask of network and inter-network regions
by thresholding the smoothed temporal mean of the intensity
\citep[cf.][]{rutten04dot2}.  We found such a threshold at
\mbox{I=1.017} by fitting the sum of two Gaussians to the intensity
histogram of the mean image.  The threshold is indicated as vertical
line in Fig.~\ref{fig:time_int} which divides the plot into two
regions, the left containing dark subfields, the right containing the
bright subfields.  The application of the threshold in combination
with a morphological operation to remove isolated pixels resulted in
mask M1 which is indicated in Fig.~\ref{fig:caimage}A, D, E and F as a
solid contour line.
%
%
Interestingly, the time scale distribution of subfields in
Fig.~\ref{fig:time_int} shows a gap between \mbox{100\,s} and
\mbox{200\,s} that separates long and short time scales (shaded area).
We therefore chose a threshold of \mbox{150\,s} to discriminate
between the fast and the slow evolving regimes (horizontal line in
Fig.~\ref{fig:time_int}).  Together with the intensity threshold the
new time scale criterion divides the plot into quadrants.  Applying
the time scale threshold to the image sequence results in a new mask
M2, which is indicated as dot-dashed contour in
Fig.~\ref{fig:caimage}F.  Mask M2 is almost completely contained
within mask M1.  The pixels that are attributed to the network in mask
M1 but not in M2 belong to subfields which contribute to the lower
right quadrant of Fig.~\ref{fig:time_int}.  This quadrant contains
subfields that are bright on average and exhibit short time scales.
Apparently, there are many structures which belong to the
inter-network even though they are brighter than average.
That includes subfields with bright \ion{Ca}{ii}\,K grains,
implying a connection between grains and inter-network pattern.
The majority of fast evolving subfields, however, is darker than
average and clearly belongs to inter-network regions (lower left
quadrant).  The barycenter of these subfields (with time scales below
\mbox{150\,s}) lies at \mbox{$\overline{\tpe{}} = 53$\,s} and a
normalized intensity of \mbox{0.953}.  The standard deviation in time
scale is \mbox{10\,s}, the minimum value is \mbox{$\tpe{, min} =
39$\,s}.
In contrast, network structures show long time scales,
some of which were longer than the duration of our data sequence, and
are on average brighter than inter-network structures.  Therefore,
they are found mostly in the upper right quadrant of
Fig.~\ref{fig:time_int}.  As expected, these persistent patches of
enhanced intensity match magnetic structures shown in the MDI
magnetogram (Fig.~\ref{fig:caimage}D) identifying them as part of the
network.  There are only few dark and slowly changing structures which
appear in the upper left quadrant.

We computed temporal power spectra for each pixel and averaged over
network and inter-network pixels according to mask M2.  The result in
Fig.~\ref{fig:power1d} shows a clearly enhanced signal in the
frequency band around \mbox{5.5\,mHz} within inter-network regions.
This can also be seen in the power map of Fig.~\ref{fig:caimage}E
where the band around \mbox{5.5\,mHz} with a band pass of
\mbox{0.9\,mHz} is displayed.  The network mask is again outlined by
contour lines.  The pixels of decreased power are almost completely
contained within M2.

\section{Discussion and Conclusion}
\label{sec:discuss}

We provide a new method to discriminate network and inter-network
structures in filtergrams which show chromospheric structure when the
data are not accompanied by information about the underlying magnetic
field.  Figure \ref{fig:time_int} demonstrates that the pattern
evolution time scale provides an excellent criterion to distinguish
between network and inter-network regions: all inter-network points are
grouped in the time interval below \mbox{100\,s}, whereas the network
has time scales above \mbox{200\,s}.  Temporal thresholding delivers
an accurate mask, M2, for network pixels.

Structures in the inter-network similar to those found in our
observations have already been detected in various other works.
\citet[][hereafter LW05]{leenaarts05} investigate a similar pattern at higher spatial
resolution -- but lower spectral resolution -- and identified it with
reversed granulation.  This result is also supported by the much
longer time scale of 2\,min they determine for the pattern.  In Fig.~1
of \citet{kneer93} the pattern can be surmised. \citet{tritschler02} present
observational data with time scales of
inter-network structures close to the ones obtained in this work.
Their spatial resolution did not suffice to resolve the small-scale
pattern and they used a broader bandwidth of 0.6\,\AA.  This work is
the first one to observe the small-scale pattern with a mean evolution
time scale much shorter than \mbox{100\,s}.

In order to verify the chromospheric origin of the detected
inter-network pattern, we compare its spatial and evolution time
scale to the recent 3D model by W04.  The temperature amplitudes of
the model chromosphere might be somewhat uncertain due to the
simplified treatment (grey, LTE), whereas the excitation of the waves occurs
in realistically modelled lower layers, making the dynamics and related time
scales more reliable.  A direct comparison with synthetic intensity maps
for \mbox{\ion{Ca}{ii}\,K$_\mathrm{2v}$} is not possible yet as the model
chromosphere is dominated by strong shocks that cause numerical
problems. We can, however, perform a qualitative comparison with
horizontal slices of gas temperature from the model chromosphere. The
cuts exhibit a pattern of granular size at a geometrical height
around \mbox{1000\,km} above \mbox{$\tau$\,=\,1} which is rapidly
evolving with time scales of around \mbox{$\tpe{} = 20$ - 30\,s}.
The pattern is quite similar to the one we observe, but the
evolution time scale is still a factor of two shorter than the
observational value.
This can be explained as follows:
\begin{figure}[tp]
 \centering
 \includegraphics[width=8.5cm]{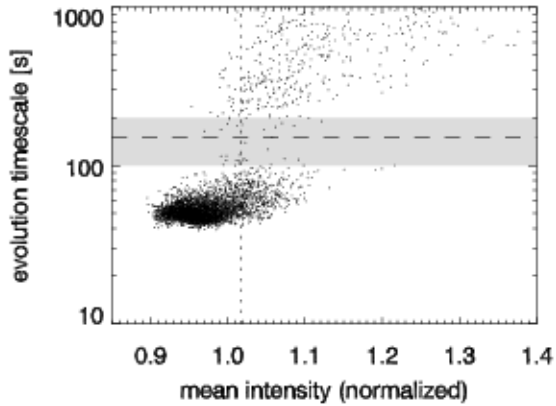}
 \caption{Time scale versus intensity. The gap between 100 and
   \mbox{200\,s} indicated with a shaded horizontal band shows the
   different behavior of network and inter-network structures.  }
 \label{fig:time_int}
\end{figure}
First of all, the correlation between gas
temperature at \mbox{$z = 1000$\,km} and intensity in the
\mbox{\ion{Ca}{ii}\,K$_\mathrm{2v}$} spectral feature might be poor as the
assumption of LTE is not valid in this strongly scattering line.  In LTE local
changes in temperature are instantaneously translated into changes
of the source function and thus intensity changes although the
intensity response is non-linear in amplitude.  In contrast to the
temperature slices at a well-defined geometrical height, the emergent
intensity (at a given wavelength) is a non-local quantity that has
contributions from an extended height range.  For the line core,
the main contribution of intensity originates from chromospheric
heights around \mbox{1000 - 1500\,km} above \mbox{$\tau$\,=\,1}
\citep[cf.][]{val81}, but already at \mbox{0.15\,\AA} off-core
considerable contributions to the filtergram intensity from lower
geometrical heights have to be expected (Rammacher, priv.
comm.).  The emergent intensity is thus integrated along
the line of sight over many layers which individually would show
different pattern evolution times \citep[decreasing with height,
see][]{wedemeyerphd}.  The integrated pattern is consequently
smoothed so that the relative intensity variations become smaller and
result in longer evolution time scales.
%
%
This effect is even enhanced due to the fact that the observed intensity
refers to corrugated surfaces of optical depth instead of plane horizontal
cuts as used for the
model.  The intensity pattern will be smoothed even more compared to
the fine structure in the temperature cuts.  W04 calculate a time
scale of $\tpe{, z} >$\,120\,s for changes in gas temperature in a
plane horizontal cut at the bottom of the photosphere, whereas the
grey emergent intensity (under assumption of LTE, which reflects
conditions at the bottom of the photosphere), yields an increase of
the typical time scale to $\tpe{, \tau} = 200$\,s.  This is
due to the fact that variations tend to be smaller on surfaces of
optical depth because these are shaped by inhomogeneities themselves.
%
In addition, seeing and instrumental effects lead to a spatial
smearing of the intensity signal and thus to longer time scales.
We found that smoothing the temperature cuts to the spatial resolution encountered
in our data increases the time scales by about 25\%.
The simulations of W04 do not include magnetic field. The existence of
inter-network fields (with intrinsic field strengths varying from
\mbox{1\,kG} down to \mbox{$\la$100\,G}) is undisputed as they have
been recently measured with high polarimetric precision
\citep[see ][and references therein]{socas+lites+pillet2004, lites+socas2004}.
The contribution of possibly unresolved magnetic elements in
inter-network regions tends to lengthen the
effective evolution time scale. A preliminary analysis of temperature cuts
from the 3D magnetohydrodynamics model by
\citet{schaffenberger05}
results in a slightly higher value of $\tpe{,MHD} = 27$\,s at the same height ($z = 1000$\,km)
compared to 24\,s in the purely hydrodynamical model by W04.
%
   \begin{figure}[tp]
   \centering
   \includegraphics[width=8.5cm]{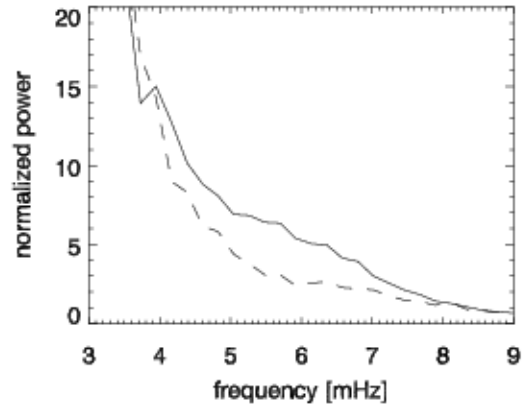}
      \caption{Temporal power for network (dashed) and inter-network (solid).}
         \label{fig:power1d}
   \end{figure}
%
Consideration of all these effects leads to a reasonable agreement
between observation and numerical model.
An interpretation of the pattern as reversed granulation is
inappropriate as the corresponding time scale is much longer than
the 53\,s found here.  LW05 state 120\,s and 90\,s for observed and
synthetic \mbox{\ion{Ca}{ii}\,H} line wing intensities,
respectively, which clearly originate from the mid-photosphere
\citep[LW05,][]{wedemeyerphd}.  These values are in line with the
time scale of order of 90\,s for the reversed granulation pattern
observed in the Fe\,I line core at 709.0\,nm by
\citet{janssen06}.
We conclude that the small-scale
inter-network pattern reported here evolves too fast to be due to
reversed granulation and is thus of chromospheric origin.  The
pattern
is very likely the intensity signature of the
mesh-like, fast evolving pattern found in recent 3D hydrodynamical
models of the solar chromosphere.  It is presumably a product of
propagation and interaction of shock waves (W04) that are excited
below the observed layer.
The same is true for the \ion{Ca}{ii} grains \citep[see
also][]{carlsson97a} that we
consider an integral part of the pattern.  Collision of two neighbouring wave
fronts results in compression and
heating of the gas at the contact region which shows up as a band of
enhanced intensity, making up the ``strings'' of the mesh-like
pattern.  At a mesh vertex, however, more than two wave fronts meet,
resulting in stronger compression and thus in an intensity higher than
the remaining pattern.  Due to the limitations of earlier
observations (see Table~\ref{tbl:cadata}) only the brightest
components have been detected as grains whereas the pattern itself
remained undetected.
A more detailed study including the simultaneous G-band data is in preparation.
\begin{acknowledgements}
We thank A.~Tritschler and R. Hammer for helpful discussions.
K.~Janssen kindly provided observational data for comparison.
F.~W\"oger acknowledges support through the {\em Landesgraduierten F\"orderungsgesetz des Landes Baden-W\"urttemberg}.
S.~Wedemeyer-B\"ohm was supported by the {\em Deutsche Forschungs\-gemein\-schaft (DFG)}, project Ste~615/5.
\end{acknowledgements}

\bibliographystyle{aa}

\end{document}